\documentstyle[12pt,a4]{article}
\begin{document}
\def\doublespaced{\baselineskip=\normalbaselineskip\multiply\baselineskip 
  by 150\divide\baselineskip by 100}
\doublespaced
\def\lsim{~{\rlap{\lower 3.5pt\hbox{$\mathchar\sim$}}\raise 1pt\hbox{$<$}}\,}
\def\gsim{~{\rlap{\lower 3.5pt\hbox{$\mathchar\sim$}}\raise 1pt\hbox{$>$}}\,}
\def\thisday{~\today ~and~ hep-ph/9810542~~}


\begin{titlepage}
\vspace{0.5cm}
\begin{flushright}
\large
hep-ph/9810542\\
July 1999  
    \hfill KEK-TH-603 \\
\end{flushright}
\vspace{0.5cm}
\begin{center}
\large
{Common Hierarchical Lepton Mass Textures for Atmospheric and Solar Neutrino 
Oscillations}
\end{center}  
\begin{center}
{\bf Ehab Malkawi$^{a,b,}$\footnote{Currently, on leave at KEK, Japan, 
e-mail:malkawi@theory.kek.jp}}
\end{center}
\begin{center}
{$^a$Department of Physics,
Jordan University of Science \& Technology\\
 Irbid 22110, Jordan}
\end{center}
\begin{center}
$^b$Theory Group, KEK, Tsukuba, Ibaraki 305-0801, Japan
\end{center} 
\vspace{0.4cm}
\raggedbottom 
\relax
\begin{abstract}
\noindent
We propose and study a common hierarchical mass texture for both 
charged lepton and neutrino sectors. The texture gives rise to a large  
$\nu_\mu\leftrightarrow\nu_\tau$ mixing. 
Also it leads to the small angle MSW effect solution for solar neutrino data.
It naturally gives a small $\nu_e\leftrightarrow \nu_\tau$ oscillation
relevant to the CHOOZ result in certain mass limits. 
A special case of only 4 input parameters leads to a consistent solution
to neutrino oscillation and lepton masses. 
\end{abstract}

\vspace{0.5cm} 
\vspace{1.0cm} 
\end{titlepage}

\section{Introduction}

Recent results from Super-Kamiokande experiment \cite{kam} 
have strengthened the evidence that atmospheric
neutrinos are depleted as they traverse the earth giving a high possibility
for an explanation in terms of neutrino oscillation. Similarly, the 
solar-neutrino anomaly as indicated by solar neutrino experiments \cite{solar} 
can be 
explained in terms of oscillation phenomena. Therefore, a hint for new 
physics beyond
the Standard Model (SM) seems to be emerging through neutrino physics.

Neutrino oscillation can explain both solar and atmospheric 
data in terms of 3-generation neutrinos  \cite{bar1,gen3} 
(ignoring the LSND results \cite{lsnd}.)
In the simplest explanation picture, solar neutrino data can be understood
in terms of $\nu_e\leftrightarrow \nu_\mu$ oscillation with a small 
mass splitting not to influence atmospheric data. 
On the other hand atmospheric data can be explained in
terms of $\nu_\mu\leftrightarrow \nu_\tau$ 
large mixing with a large 
mass splitting compared to the $\nu_e\leftrightarrow\nu_\mu$ case \cite{bar1}.

Hints about correlated mass splittings and mixing angles are a relevant guide 
in constructing lepton mass matrices.
Search for lepton mass textures can also be guided by the quark sector which 
exhibits a definite hierarchical structure. 
However, as suggested by atmospheric data,   
large mixing between the $\nu_\mu$ and $\nu_\tau$ neutrinos implies 
different behavior from the usual quark sector. 
Nevertheless, it remains an appealing approach if we can understand solar and 
atmospheric data in terms of  hierarchical
lepton mass matrices. 
Hierarchical structures are attractive because they reflect 
the correspondence between the quark and lepton sectors.
This correspondence can form an underground for 
a grand unified picture in understanding the fermion mass generation.
Several authors have considered this possibility proposing several  
lepton mass textures \cite{xing}. 

In this work we suggest to explain both solar and atmospheric neutrino data
in terms of a common hierarchical lepton mass matrix different than
those already considered. We treat both charged lepton and neutrino sectors
equally by using a common mass texture.   
With this proposed mass matrix we can explain 
the apparent large $\nu_\mu\leftrightarrow\nu_\tau$ mixing angle. 
It naturally leads to 
the small angle MSW effect solution \cite{msw} to the solar neutrino problem. 
In addition, it naturally gives
a small mixing angle between $\nu_e$ and $\nu_\tau$ which can be relevant
as suggested  by the CHOOZ experiment in certain mass limits \cite{chooz}. 

Recent analysis of  all solar neutrino data \cite{bar1,fogli} indicates 
that for the small angle MSW effect solution 
\begin{equation}
\Delta m^2_{\rm{sun}}\approx (0.3-1.2)\times 10^{-5}\,\, {eV^2}\, ,
\end{equation}
\begin{equation}
\sin^2 2\theta_{\rm{sun}}=(0.3-1.2)\times 10^{-2}\, .
\end{equation}
While for the large angle solution 
\begin{equation}
\Delta m^2_{\rm{sun}}\approx (0.8-3.0)\times 10^{-5}\,\, {eV^2}\, ,
\end{equation}
\begin{equation}
\sin^2 2\theta_{\rm{sun}} \approx 0.42-0.74\, .
\end{equation}
Solar anomaly can also be explained by vacuum oscillation if \cite{fogli}
\begin{equation}
\Delta m^2_{\rm{sun}}\approx (0.6-1.1)\times 10^{-10}\,\, {eV^2}\, ,
\end{equation}
\begin{equation}
\sin^2 2\theta_{\rm{sun}}\geq 0.70\, .
\end{equation}
The atmospheric anomaly can be explained by $\nu_\mu\leftrightarrow\nu_\tau$
oscillation if \cite{kam,bar1,barger}
\begin{equation}
\Delta m^2_{\rm{atm}}\approx (0.3 - 8)\times 10^{-3}\,\, {eV^2}\, ,
\end{equation}
\begin{equation}
\sin^2 2\theta_{\rm{atm}}\geq 0.75\, .
\end{equation}

Certainly, the above results are very intriguing but are not sufficient 
to uniquely determine lepton mass matrices. This is true because, 
in principle, one needs to construct
the charged lepton and the $6\times 6$ neutrino mass matrices, 
if neutrino masses 
are realized via the see-saw mechanism. Different assumed symmetries, 
continuous or discrete, can be imposed with different types of 
symmetry-breaking scenarios.

\section{Hierarchical lepton masses}

Since both charged lepton and neutrino mixing angles are required to build
the mixing matrix $V=V_e^\dagger V_\nu$ a construction of the 
lepton mass matrix is not straightforward.
Usually, one needs to start from an assumed structure, which may be guided
by some proposed symmetry, for the charged lepton and  
neutrino sectors, then working out the correspondence between different 
quantities. 
In this work we pursue the main assumption that a common mass structure holds  
for both 
charged lepton and neutrino sectors. 
Hence, we treat the two sectors on an equal footing.
This assumption usually is not what what we expect from famous grand unified
theories. Therefore, if this assumption holds then it can give a new hint for 
the type of underlying physics generating this behavior.

We work with the following parameterization of the lepton flavor mixing matrix
$V=R_{23}(\theta_{23})R_{13}(\theta_{13})R_{12}(\theta_{12})$.
Explicitly, $V$ is written as
\begin{equation}
V=\pmatrix{c_{13}c_{12} &c_{13}s_{12}  & s_{13} \cr 
-c_{23}s_{12}-s_{23}s_{13}c_{12} & c_{23}c_{12}-s_{23}s_{13}s_{12} & 
c_{13}s_{23} \cr s_{23}s_{12}-c_{12}c_{23}s_{13} &
 -c_{12}s_{23}-c_{23}s_{12}s_{13} & c_{13}c_{23} \cr}\, ,
\end{equation}
where 
$R_{ij}(\theta_{ij})$ represents a rotation in the $ij$ plane by an angle
$\theta_{ij}$, and where 
$s_{ij}\equiv \sin\theta_{ij}$, $c_{ij}\equiv \cos\theta_{ij}$. 
The mixing angles are taken 
$0\leq \theta_{ij}\leq \pi/2$. The phases have been dropped as we, 
for simplicity, work with real mass matrices.

We stress that we want to express both neutrino and charged lepton 
masses using the same matrix form. 
We also realize the neutrino mass matrix via the
see-saw mechanism. 
In constructing the mass matrix of the charged and neutral leptons 
we make the following two assumptions.
\begin{itemize}
\item[1.]
The tree-level mass matrix has a few number of parameters, of the
same order, reflecting  
a non trivial symmetry. The proposed mass matrix provides, 
as shown later,  
a phenomenological solution 
to the neutrino oscillation and controls both 
charged and neutral lepton masses. Therefore, it is highly desirable to 
understand the symmetry which governs the structure of the mass matrix.
It turns out that the symmetry of the proposed mass structure 
involves a non trivial rotation of the flavor
states in a complex space.   
A realization of the symmetry is given later when we introduce explicitly 
the mass matrix form.    
However, the symmetry has to be broken in order to give the suitable 
mass spectrum. 
The breaking is expressed in terms of a small parameter $\epsilon$ for both
charged and neutral lepton mass matrix. 
Thus, the physical mass matrix can be written as  a perturbative
series in powers of the small parameter $\epsilon$. 
\item[2.]
The proposed symmetry is not enough to guarantee the hierarchical 
structure of the lepton masses. Therefore, we make an extra assumption
in the construction of the mass matrix. 
We require the tree-level charged and neutral lepton masses 
to satisfy the condition $m_{1,2}=0$ while $m_3\neq 0$, 
where $m_i$ refers to the mass of the $i$th generation charged and neutral 
lepton. 
This requirement reduces further 
the number of parameters of the mass matrix by one.
Due to the symmetry breaking, lepton masses are generated with 
a hierarchical structure, i.e., $m_1=O(\epsilon^2)$, $m_2=O(\epsilon)$, 
while $m_3=O(1)$. Hence, we conclude that 
the symmetry-breaking parameter  $\epsilon \sim m_2/m_3 $\, .
\end{itemize}

Starting with the charged lepton sector, we propose the following
tree-level mass matrix
\begin{equation}
M_E=\lambda v\pmatrix{CD & C & C \cr D & 1 & 1 \cr D & 1 & 1 \cr}\, ,
\label{eq1}
\end{equation}
where C and D are taken of order 1 and where left (right) handed leptons 
are contracted to the left (right), $\overline{e_L} M_E e_R$.
The symmetry responsible for this structure can be realized for the
left-handed sector by the 
following discrete rotation
\begin{equation}
\pmatrix{e_L \cr \mu_L \cr \tau_L \cr} 
\rightarrow \frac{1}{2} 
\pmatrix{2 & 0 & 0 \cr 0& 1+i & 1-i \cr 0 & 1-i & 1+i \cr} 
\pmatrix{e_L \cr \mu_L \cr \tau_L} \, .
\end{equation} 
While for the right handed sector we have
\begin{equation}
\pmatrix{e_R \cr \mu_R \cr \tau_R \cr} 
\rightarrow \frac{1}{2} 
\pmatrix{2 & 0 & 0 \cr 0& 1-i & 1+i \cr 0 & 1+i & 1-i \cr} 
\pmatrix{e_R \cr \mu_R \cr \tau_R} \, .
\end{equation} 
It is easy to verify that $e_{L,R}$ and ${(\mu+\tau)}_{L,R}$ are 
invariant under the above transformation while ${(\mu-\tau)}_{L,R}$ is not.
The same transformation is assumed to hold for the neutrino sector. 
Note that $M_{E11}=CD$ is imposed in order to guarantee the second 
assumption, namely $m_e=m_\mu=0$ at this order. 
The symmetry of the lepton mass matrix can be easily generalized to 
a continuous complex rotation, even though we do not pursue it in this
work.
The mass matrix $M_E$ can be rewritten as 
\begin{equation}
M_E=\lambda v \pmatrix{C & 0 &0 \cr 0 & 1 & 0 \cr 0 & 0 & 1 \cr}
\pmatrix{1 & 1 & 1 \cr 1 & 1 & 1 \cr 1 & 1 & 1 \cr}
 \pmatrix{D & 0 &0 \cr 0 & 1 & 0 \cr 0 & 0 & 1 \cr} \, .
\end{equation}
This texture can be understood
if C is interpreted as the relative mass coupling of the left-handed electron
with respect to the left-handed muon or tau. 
Similarly, $D$ is interpreted as the 
relative mass coupling of the right-handed electron
with respect to the right-handed muon or tau.
For the special case $C=D=1$, the matrix reduces to the exact flavor 
democratic structure which is discussed in section 3.
At the tree level, the masses are 
$m_e=m_\mu=0$ and $m_\tau=\lambda v\sqrt{(C^2+2)(D^2+2)}$.
The relevant charged lepton mixing angles are 
$\theta^e_{23}=\pi/4$, $\tan\theta^e_{13}=C/\sqrt{2}$, while
$\theta^e_{12}$ is arbitrary at this stage. However, once we break 
the symmetry we find that $\theta^e_{12}$ is very close to $\pi/2$.

Next, we consider the right-handed neutrino sector. At the zeroth order 
of $\epsilon$ the mass matrix is given by a similar form to the charged
lepton mass matrix,
\begin{equation}
M_{RR}=M\pmatrix{A^2 & A & A \cr A & 1 & 1 \cr A & 1 & 1 \cr}\, ,
\end{equation}
where $A$ is of order 1 and can be interpreted as the relative 
mass coupling of $\nu_{eR}$ with respect 
to the $\nu_{{\mu,\tau} R}$ neutrinos.
At this order, only one right-handed neutrino is massive with  a mass 
$M_3=M(A^2+2)$. However, once we break the symmetry other right-handed 
neutrinos acquire their hierarchical masses. 
Therefore, the lightest state will have a mass of order 
\begin{equation}
M_1\approx M_3\, {(\frac{m_\mu}{m_\tau})}^2 \approx 0.003\,M_3\, , 
\end{equation}
which is relatively heavy if $M$ is taken of the order of the GUT 
scale. 

For the Dirac part of the neutrino matrix we have
\begin{equation}
M_{LR}=\lambda^\prime v\pmatrix{AB & B & B \cr A & 1 & 1 \cr A & 1 & 1 \cr}
\, ,
\end{equation}
where $B$ is of order 1.
Notice that with the interpretation of $A$ as the relative mass coupling 
of $\nu_{eR}$, we are forced to include it
in $M_{LR}$. Thus the matrix $M_{LR}$ contains two additional 
parameters $B$ and $\lambda^\prime$. The parameter $B$ 
is given the interpretation as the relative mass coupling
of the left-handed electron neutrino, $\nu_{eL}$, compared to 
the left-handed muon or tau neutrinos.
The general $6\times 6$ neutrino mass matrix is then given by 
\begin{equation}
M_\nu= \pmatrix{0 & M_{LR} \cr M_{LR}^{Tr} & M_{RR} \cr}\, .
\end{equation}
The general mass matrix is singular and the see-saw mechanism 
at this stage leads, as shown later, to
only a one massive left-handed neutrino.  
The symmetry breaking is responsible
for generating masses to all right-handed states which then give
mass to the light states via the see-saw mechanism.

We next introduce a small 
perturbation $\epsilon$ which breaks down the universality between 
the second and third generation. At the first order in $\epsilon$ 
we simply require the mass coupling 
of the third generation lepton to deviate from the second generation 
lepton by $\epsilon$.  
For the charged lepton case we write the mass matrix in the form 
\begin{equation}
M_E=\lambda v\pmatrix{C & 0 & 0 \cr 0 & 1 & 0 \cr 0 & 0 & 1+\epsilon\cr}
\pmatrix{1+\kappa \epsilon & 1 & 1 \cr 1 & 1 & 1 \cr 1 & 1 & 1\cr}
\pmatrix{D & 0 & 0 \cr 0 & 1 & 0 \cr 0 & 0 & 1+\epsilon\cr}
\, ,
\end{equation}
where $\kappa$ is a constant of order 1 introduced to allow a different 
contribution in the element $M_{E_{11}}$ which is evidently not related
to other elements by symmetry. 
We do not modify $C$ and $D$ by a factor of $\epsilon$ 
because we can then simply 
redefine those parameters to absorb the new factor. 
Note that this structure guarantees that
the masses generated are of hierarchical nature as shown below. 
The new mass matrix can be rewritten, up to the order of $\epsilon$, as
\begin{equation}
M_E=\lambda v\pmatrix{CD(1+\kappa \epsilon) & C & C(1+\epsilon) 
\cr D & 1 & 1+\epsilon \cr D(1+\epsilon) & 1+\epsilon & 
1+2\epsilon \cr}\, .
\end{equation}
The physical masses are hierarchical and given, to leading order, as 
\begin{equation}
m_e=O(\epsilon^2)\, ,
\end{equation}
\begin{equation}
m_\mu=\lambda v\frac{2\kappa CD}{\sqrt{(C^2+2)(D^2+2)}} \epsilon \, ,
\end{equation}
and
\begin{equation}
m_\tau=\lambda v{\sqrt{(C^2+2)(D^2+2)}+O(\epsilon)}\, .
\end{equation}
Thus, the small perturbation $\epsilon$ can be written as
\begin{equation}
\epsilon = \frac{(C^2+2)(D^2+2)}{2\kappa CD} \frac{m_\mu}{m_\tau}\, .
\end{equation}
It is easy to verify that there is a lower bound on $\epsilon$ given as 
$\epsilon \gsim \frac{0.23}{\kappa}$. Since $\kappa$ is of order 1, 
the breaking parameter $\epsilon$ can not be very small.
The charged lepton mixing angles, to leading order, are 
\begin{equation}
\tan\theta^e_{23}=1-\epsilon\, ,
\end{equation}
\begin{equation}
\tan\theta^e_{13}=\frac{C}{\sqrt{2}}-\frac{C}{\sqrt{2}}
\left(\frac{1}{2}-\frac{D^2}{D^2+2}\right)\epsilon \, ,
\end{equation}
while $\theta^e_{12}$ is close to $\pi/2$,
\begin{equation}
\tan\theta^e_{12}=\frac{4\kappa C}{\sqrt{(C^2+2)}\epsilon}+O(1)\, .
\end{equation}

For the case of $M_{RR}$, we similarly introduce the same small  
perturbation $\epsilon$. The mass matrix in this case is   
\begin{equation}
M_{RR}=M\pmatrix{A^2(1+\kappa\epsilon) & A & 
A(1+\epsilon) \cr A & 1 & 1+\epsilon \cr A(1+\epsilon) & 
1+\epsilon & 1+2\epsilon \cr}\, .
\end{equation} 
The matrix $M_{RR}$ has a hierarchical structure which predicts a 
heavy right-handed neutrino $M_3=O(1)$ and two relatively lighter ones,
$M_2=O(\epsilon)$ and $M_1=O(\epsilon^2)$. However, as we found that 
$\epsilon$ can not be a very small parameter, all mass
states are relatively heavy, taking $M$ to be at the GUT scale.

Introducing the small perturbation $\epsilon$ into $M_{LR}$ 
similar to $M_E$ and $M_{RR}$ we find
\begin{equation}
M_{LR}=\lambda^\prime v\pmatrix{AB(1+\kappa\epsilon) & B & 
B(1+\epsilon) \cr A & 1 & 1+\epsilon \cr A(1+\epsilon) & 
1+\epsilon & 1+2\epsilon \cr}\, .
\end{equation}
The light neutrino mass matrix is then given via the see-saw mechanism,
$M_{LL}=M_{LR}M^{-1}_{RR}M^{T}_{LR}$. We find that
\begin{equation}
M_{LL}=\frac{{(\lambda^\prime v)}^2}{M}
\pmatrix{B^2(1+\kappa\epsilon) & B & 
B(1+\epsilon) \cr B & 1 & 1+\epsilon \cr B(1+\epsilon) & 
1+\epsilon & 1+2\epsilon \cr}\, .
\end{equation} 
Therefore, the parameters we introduce to account for the low energy 
physics are: $\lambda$, $\lambda^\prime$, $B$, $C$, $D$, $\epsilon$, 
and $\kappa$ which amount to 7 parameters needed to account for 
6 masses and 3 mixing angles, ignoring phases. Thus, two predictions
of the model are expected. However, it is interesting that the special case
discussed in section 3 with only 4 parameters leads also to 
a consistent result.   
Note that the parameter $A$ has dropped out in our final result. 
Also note that we could start studying $M_{LL}$ from an effective 
point of view without adhering to the see-saw mechanism.
The light neutrino masses are found to be 
\begin{equation}
m_1=O(\epsilon^2)\, ,
\end{equation}
\begin{equation}
m_2=\frac{{(\lambda^\prime v)}^2}{M}\frac{2B^2\kappa}{(B^2+2)} 
\epsilon\, ,
\end{equation}
and
\begin{equation}
m_3=\frac{{(\lambda^\prime v)}^2}{M}(B^2+2) +O(\epsilon)\, . 
\end{equation}
Thus, the ratio
\begin{equation}
\frac{m_2}{m_3}=\frac{B^2}{CD}\frac{(C^2+2)(D^2+2)}{{(B^2+2)}^2} 
\frac{m_\mu}{m_\tau}\approx \frac{m_\mu}{m_\tau}= 0.06\, , 
\end{equation}
where we took $B=C=D\approx 1$.
The relevant mass differences, to leading order, are
\begin{equation}
\Delta m^2_{21}=m_2^2-m_1^2=\frac{{(\lambda^\prime v)}^4}{M^2}\,
\frac{4B^4{\kappa}^2}{{(B^2+2)}^2}\epsilon^2 \, ,
\end{equation}
\begin{equation}
\Delta m^2_{32}=m_3^2-m_2^2=\frac{{(\lambda^\prime v)}^4}{M^2}\, 
(B^2+2)^2 +O(\epsilon)\, .
\end{equation}
The neutrino mixing angles are
\begin{equation}
\tan\theta^\nu_{23}=1-\epsilon \, , 
\end{equation}
\begin{equation}
\tan\theta^\nu_{13}=\frac{B}{\sqrt{2}}- \frac{B}{\sqrt{2}}\left(\frac{1}{2}-
\frac{B^2\kappa}{B^2+2}\right)\epsilon \, .
\end{equation}
while $\theta^\nu_{12}$ is very close to $\pi/2$,
\begin{equation}
\tan\theta^\nu_{12}=\frac{4C\kappa}{\sqrt{(B^2+2)}\epsilon}+
O(1)\, .
\end{equation}

The combined mixing matrix defined as $V=V_e^\dagger V_\nu$
is constructed. We find that, to leading order, 
$\theta_{23}=\theta^\nu_{13}-\theta^e_{13}$ and that
\begin{equation}
\tan\theta_{23}=\frac{\sqrt{2}(B-C)}{2+BC}+O(\epsilon)\, ,
\end{equation}
\begin{equation}
\tan\theta_{13}=O(\epsilon)\, ,
\end{equation}
\begin{equation}
\tan\theta_{12}=O(\epsilon)\, .
\end{equation}
Therefore, we are lead naturally to a large mixing angle $\theta_{23}$ and 
small $\theta_{12}$ and $\theta_{13}$.
To leading order the mixing matrix $V$ can be written as
\begin{equation}
V=\pmatrix{1 & O(\epsilon) & O(\epsilon) \cr
O(\epsilon) & \frac{BC+2}{\sqrt{(C^2+2)(B^2+2)}}& 
\frac{\sqrt{2}(B-C)}{\sqrt{(C^2+2)(B^2+2)}}\cr
 O(\epsilon)& -\frac{\sqrt{2}(B-C)}{\sqrt{(C^2+2)(B^2+2)}}&
\frac{BC+2}{\sqrt{(C^2+2)(B^2+2)}}\cr} \, .
\end{equation}

\section{Special Case}

In this section we consider the interesting special case
 where $C=D=1$. This is the case where exact flavor democracy
is assumed in the charged lepton sector. 
We also take $\kappa=1$ and thus leave
4 free input parameters; $\lambda$, $\lambda^\prime$, $B$, and $\epsilon$
to be determined from data. The mass ratio 
$m_\mu/m_\tau=2/9\,\epsilon$ and hence we conclude 
$\epsilon=4.5\, m_\mu/m_\tau \approx 0.25$. Therefore, we find 
\begin{equation}
\sin^2 2\theta_{23}=\frac{8}{9}{\left(\frac{B^2+B-2}{B^2+2}\right)}^2
+O(\epsilon)\, .
\end{equation}
To satisfy the condition $\sin^2 2\theta_{23}\gsim 0.75$, $B$ is required to
be larger than 3.
For large $B$, $\sin^2 2\theta_{23}\approx 8/9$ independent of masses.
The relevant neutrino mass ratio is  
\begin{equation}
\frac{\Delta m^2_{12}}{\Delta m^2_{23}}\approx \frac{4}{B^2}\epsilon^2 
\leq 3\times 10^{-2}\, ,
\end{equation}
which is consistent with data, and where we 
considered $B>3$ and $\epsilon\approx 0.25$. 
For $\theta_{12}$ we find that for $B\approx 3$, 
 $\sin^2 2\theta_{12}\approx 0.7\times 10^{-2}$ which is also 
consistent with data.
Thus the considered texture can give all desired outcome without fine tuning
of parameters. 
It naturally gives rise to small values of $\theta_{13}$. For 
example for the considered special case we find that, for large $B$, 
$\sin\theta_{13}\approx \sqrt{2}/4 \,\epsilon\approx 9\times 10^{-2}$, 
which is consistent
with the result of the CHOOZ experiment if $\Delta m^2_{23}> 10^{-3}\, eV^2$.
For the given range of $\epsilon\approx 0.2-0.3$, the right-handed 
neutrino states remain relatively heavy.   

Notice that the case where $B \approx C$ gives rise to a small $\theta_{23}$
which is  not
consistent with atmospheric data. Therefore, a large splitting between the 
left-handed charged lepton and neutrino parameters $B$ and $C$ 
is required to produce large $\theta_{23}$. 

In conclusion we investigated a common mass texture for both charged and 
neutral lepton sectors. The texture gives rise to all desired features 
needed to explain the solar and atmospheric data. 
It leads naturally to the small 
angle MSW effect solution 
and gives a small $\theta_{13}$ consistent with the CHOOZ result in 
certain mass limits. A special case with only 4 parameters also gives 
a consistent solution to the neutrino oscillation and controls both
charged and neutral lepton masses. 

\vspace{0.3cm}
\noindent
{\bf Acknowledgments}

\noindent
E.M. would like to thank the Matsumae International Foundation for 
offering him the Fellowship. He thanks the kind hospitality of KEK.
He also thanks K.S. Babu for useful discussion and comments. 
Part of this work is supported by Jordan University of Science \& 
Technology under grant \#77/99.

\end{document}